\documentclass[prb,twocolumn,showpacs,amsmath,superscriptaddress,psfig,amssymb]{revtex4-1}
\usepackage{amsmath}
\usepackage{amssymb}
\usepackage{graphicx}
\usepackage{dcolumn}
\usepackage{bm}
\usepackage{txfonts}
\usepackage{tabularx}
\usepackage{cases}

\begin{document}
\title{Nodeless superconductivity and the peak effect in the quasi-skutterudites $\mathrm{Lu}_3\mathrm{Os}_4\mathrm{Ge}_{13}$ and $\mathrm{Y}_3\mathrm{Ru}_4\mathrm{Ge}_{13}$}

\author{Z.F. Weng}
\affiliation{Center for Correlated Matter and Department of Physics, Zhejiang University, Hangzhou 310058, China}
\author{M. Smidman}
\email{msmidman@zju.edu.cn}
\affiliation{Center for Correlated Matter and Department of Physics, Zhejiang University, Hangzhou 310058, China}
\author{G.M. Pang}
\affiliation{Center for Correlated Matter and Department of Physics, Zhejiang University, Hangzhou 310058, China}
\author{O. Prakash}
\affiliation{Department of Condensed Matter Physics and Materials Science, Tata Institute of Fundamental Research, Mumbai 400005, India}
\author{Y. Chen}
\affiliation{Center for Correlated Matter and Department of Physics, Zhejiang University, Hangzhou 310058, China}
\author{Y.J. Zhang}
\affiliation{Center for Correlated Matter and Department of Physics, Zhejiang University, Hangzhou 310058, China}
\author{S. Ramakrishnan}
\affiliation{Department of Condensed Matter Physics and Materials Science, Tata Institute of Fundamental Research, Mumbai 400005, India}
\author{H. Q. Yuan}
\email{hqyuan@zju.edu.cn}
\affiliation{Center for Correlated Matter and Department of Physics, Zhejiang University, Hangzhou 310058, China}
\affiliation{Collaborative Innovation Center of Advanced Microstructures, Nanjing 210093, China}
\date{\today}

\begin{abstract}
We report an investigation of the superconducting states of $\mathrm{Lu}_3\mathrm{Os}_4\mathrm{Ge}_{13}$ and $\mathrm{Y}_3\mathrm{Ru}_4\mathrm{Ge}_{13}$ single crystals by  measurements of the electrical resistivity, ac susceptibility and London penetration depth. The analysis of the  penetration depth and the derived superfluid density indicates the presence of nodeless superconductivity and suggest that there are multiple superconducting gaps in both materials. Furthermore, ac susceptibility measurements of both compounds display the peak effect in the low temperature region of the $H-T$ phase diagram. This anomalous increase of the critical current with field gives an indication of a change of the arrangement of flux lines in the mixed state, as found in some of the isostructural stannide materials.
\end{abstract}

\pacs{74.25.Ha, 74.25.Uv, 74.25.Bt}
\maketitle

\section{introduction}
Since the discovery of the ternary superconducting stannides $R_3T_4M_{13}$ \cite{REMEIKA1980923,HODEAU1980839}, numerous compounds with this stoichiometry have been synthesized, where $R$ is an alkaline earth or rare earth metal, $T$ is a transition metal, and $M$ is one of In, Si, Ge, Sn or Pb. At room temperature, most of these materials crystallize in the primitive cubic $\mathrm{Yb}_3\mathrm{Rh}_4\mathrm{Sn}_{13}$ type structure with the space group $Pm\bar{3}n$ \cite{HODEAU1980839}. In this caged structure, $R$ and $T$ atoms each occupy one crystallographic site, while there are two sites for $M$, where the  $M$ atom on one site is surrounded by a polyhedral cage formed from the $M$ atoms on the other site. There are a few examples of compositions with  different structures, such as tetragonal $\mathrm{Yb}_3\mathrm{Pt}_4\mathrm{Ge}_{13}$ \cite{C2DT30339F}, monoclinic $\mathrm{Y}_3\mathrm{Pt}_4\mathrm{Ge}_{13}$ \cite{PhysRevB.87.224502} and $\mathrm{U}_3\mathrm{Ir}_4\mathrm{Ge}_{13}$, which has a  noncentrosymmetric rhombohedral structure\cite{PhysRevB.91.094110}. The $R_3T_4M_{13}$ compounds have attracted considerable interest since they display a wide range of physical properties, such as superconductivity, \cite{REMEIKA1980923,Hayamizu2011,PhysRevB.83.184509,PhysRevB.89.125111,PhysRevB.48.10435,PhysRevB.87.224502,0953-2048-28-11-115012,Prakash201390,strydom_superconductivity_2014,doi:10.1021/cm504658h,0953-8984-13-33-319} magnetism, \cite{Sato1993630,doi:10.1143/JPSJ.75.044710,PhysRevB.91.094110,PhysRevB.93.064427,PhysRevB.65.024401} mixed valence behavior, \cite{0953-8984-13-33-319,Strydom2008746,rai_intermediate_2015,PhysRevB.93.035101} structural phase transitions and quantum criticality. \cite{PhysRevLett.109.237008,PhysRevLett.114.097002}

Particular attention has been paid to the stannides  $R_3T_4\mathrm{Sn}_{13}$. $\mathrm{Sr}_3\mathrm{Rh}_4\mathrm{Sn}_{13}$ and $\mathrm{Sr}_3\mathrm{Ir}_4\mathrm{Sn}_{13}$ exhibit both a second-order structural phase transition and superconductivity, where  the structural phase transition can be  suppressed to lower temperatures by applying pressure or doping \cite{PhysRevLett.109.237008, PhysRevLett.114.097002}. Non-Fermi liquid behavior is observed when the structural phase transition temperature is tuned to zero, implying the existence of a structural quantum critical point. Evidence for strongly coupled nodeless superconductivity has been found in the stannides $R_3T_4\mathrm{Sn}_{13}$ ($R$=La, Sr, Ca and $T$=Rh, Ir) from various measurements, \cite{Hayamizu2011,PhysRevB.83.184509,PhysRevB.86.024522,PhysRevB.86.064504,PhysRevB.88.104505} where a strong enhancement of  the coupling is found in the vicinity of the structural quantum critical point in $(\mathrm{Ca}_x\mathrm{Sr}_{1-x})_3\mathrm{Rh}_4\mathrm{Sn}_{13}$ from the specific heat \cite{PhysRevLett.115.207003} and in $\mathrm{Ca}_3\mathrm{Ir}_4\mathrm{Sn}_{13}$ from $\mu$SR \cite{PhysRevB.92.195122}.

In comparison to the stannides, germanides with $M$=Ge which show a lack of a structural transition have received less attention.
In many cases weak semiconducting behavior in the resistivity is observed and those materials which are superconductors have fairly low values of $T_c$. \cite{PhysRevB.48.10435} The recent synthesis of high quality single crystals of the germanide superconductors $\mathrm{Lu}_3\mathrm{Os}_4\mathrm{Ge}_{13}$ \cite{0953-2048-28-11-115012} and $\mathrm{Y}_3\mathrm{Ru}_4\mathrm{Ge}_{13}$ \cite{Prakash201390} offers  new opportunities to investigate the superconducting properties of  the $R_3T_4\mathrm{Ge}_{13}$ series. Multiband superconductivity was suggested in  $\mathrm{Lu}_3\mathrm{Os}_4\mathrm{Ge}_{13}$ on the basis of specific heat measurements, which is consistent with the calculated complex Fermi surface where the density of states predominantly consists of contributions from Os and Ge.  \cite{0953-2048-28-11-115012} Meanwhile although  $\mathrm{Y}_3\mathrm{Ru}_4\mathrm{Ge}_{13}$ is metallic, band structure calculations suggest a broad minimum in the density of states in the vicinity of the Fermi level, which is mainly from Ge with little contribution from Ru $d$-orbitals. \cite{pavan2011exploring} However, further characterization of the superconducting order parameter of both compounds is necessary. Here, we present measurements of the resistivity ($\rho$), ac susceptibility ($\chi$) and change of the  penetration depth  [$\Delta\lambda(T)$] of $\mathrm{Lu}_3\mathrm{Os}_4\mathrm{Ge}_{13}$ and $\mathrm{Y}_3\mathrm{Ru}_4\mathrm{Ge}_{13}$ single crystals down to 0.4~K. The  penetration depth of both materials flattens  at low temperatures, indicating fully gapped superconductivity and the calculated superfluid density $\rho_s(T)$ is well described by a two-gap $s$-wave model, indicating that $\mathrm{Lu}_3\mathrm{Os}_4\mathrm{Ge}_{13}$ and $\mathrm{Y}_3\mathrm{Ru}_4\mathrm{Ge}_{13}$ are nodeless, multiband superconductors. Furthermore, field dependent ac susceptibility measurements at low temperatures show the presence of the peak effect in both compounds, which may indicate a change in the arrangement of vortices in the mixed state, as seen in some of the isostructural stannide materials.

\section{Experimental details}
$\mathrm{Lu}_3\mathrm{Os}_4\mathrm{Ge}_{13}$ and $\mathrm{Y}_3\mathrm{Ru}_4\mathrm{Ge}_{13}$ single crystals were grown using the Czochralski method in a tetra-arc furnace under an argon atmosphere, as described previously \cite{0953-2048-28-11-115012,Prakash201390}. The electrical resistivity was measured in a $^3$He cryostat utilizing a four probe method. The ac susceptibility was measured in the same $^3$He cryostat using an ac susceptometer. Note that the excitation current used was 100~$\mu$A, which corresponds to a magnetic field of around 0.4~Oe. Precise measurements of the penetration depth change $\Delta \lambda(T)$ were performed using a tunnel-diode oscillator (TDO) based, self-inductive technique at an operating frequency of 7 MHz down to 0.4 K in a $^3$He cryostat, with which a noise level as low as 0.1~Hz can be obtained. The London penetration depth change is proportional to the change of the resonant frequency $\Delta f(T)$, i.e., $\Delta\lambda(T) = \lambda(T) - \lambda(0) = G \Delta f(T)$, where $\lambda(0)$ is the penetration depth at zero temperature and the $G$  is calculated using the sample and coil geometries \cite{Gfactor}. The coil of the oscillator generates a very small ac magnetic field ($H_{ac}$ $\approx$ 20 mOe), which is much less than the lower critical fields of  $\mathrm{Lu}_3\mathrm{Os}_4\mathrm{Ge}_{13}$ and $\mathrm{Y}_3\mathrm{Ru}_4\mathrm{Ge}_{13}$ \cite{0953-2048-28-11-115012,Prakash201390} ensuring that all the measurements were performed in the Meissner state.

\section{experimental results and discussion}

\begin{figure}[t]
\includegraphics[width=0.4\textwidth]{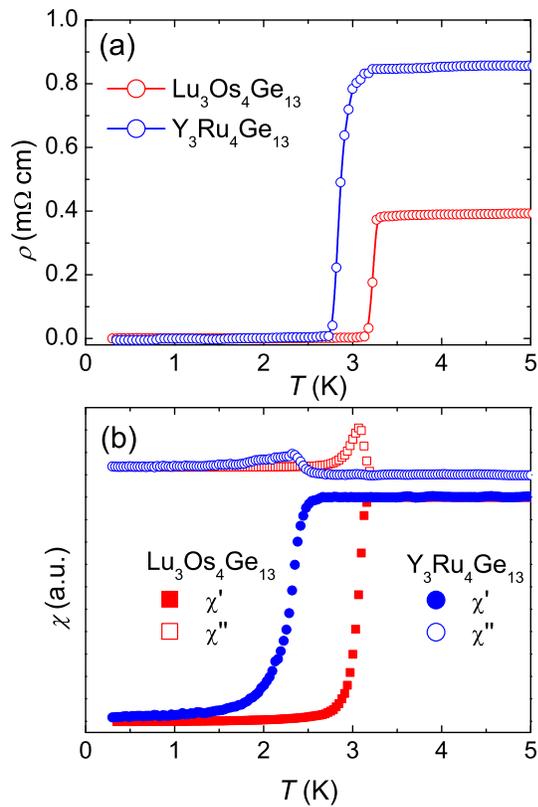}
\caption{(Color online) Temperature dependence of (a) electrical resistivity, and (b) ac susceptibility of $\mathrm{Lu}_3\mathrm{Os}_4\mathrm{Ge}_{13}$ and $\mathrm{Y}_3\mathrm{Ru}_4\mathrm{Ge}_{13}$.}
\label{fig1}
\end{figure}

In order to characterize the samples, the temperature dependence of the electrical resistivity $\rho(T)$ and ac magnetic susceptibility $\chi(T)$ for $\mathrm{Lu}_3\mathrm{Os}_4\mathrm{Ge}_{13}$ and $\mathrm{Y}_3\mathrm{Ru}_4\mathrm{Ge}_{13}$ were measured, as shown in Fig. \ref{fig1}. Superconducting transitions are observed in the measurements for  both compounds, with $T_c=3.2$~K and 3.05~K  from  midpoints of the transitions in the resistivity, and ac susceptibility of $\mathrm{Lu}_3\mathrm{Os}_4\mathrm{Ge}_{13}$,  respectively, while the respective values for  $\mathrm{Y}_3\mathrm{Ru}_4\mathrm{Ge}_{13}$  are $T_c=2.8$~K and 2.3~K. Since the  susceptibility is a bulk probe, the values of $T_c$ from the ac susceptibility are used in the later analysis of the superfluid density.

\subsection{Penetration depth and superfluid density}

\begin{figure}[t]
\includegraphics[width=0.4\textwidth]{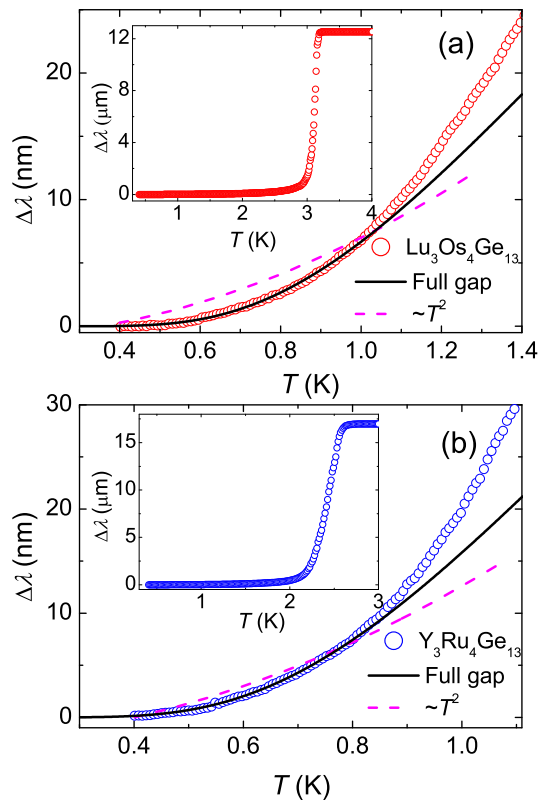}
\caption{(Color online) Temperature dependence of the change of the London penetration depth $\Delta \lambda(T)$ at low temperatures of  (a) $\mathrm{Lu}_3\mathrm{Os}_4\mathrm{Ge}_{13}$ and  (b) $\mathrm{Y}_3\mathrm{Ru}_4\mathrm{Ge}_{13}$. The solid and dashed lines show fits of $\Delta \lambda(T)$ to a fully gapped model and a $T^2$ dependence respectively.  The insets show $\Delta \lambda(T)$  from the base temperature to above $T_c$.}
\label{fig2}
\end{figure}

In Fig.~\ref{fig2}  the temperature dependence of $\Delta\lambda(T)$ is shown for (a) $\mathrm{Lu}_3\mathrm{Os}_4\mathrm{Ge}_{13}$ and (b) $\mathrm{Y}_3\mathrm{Ru}_4\mathrm{Ge}_{13}$, which  are converted from the frequency shift $\Delta f(T)$ with respective calibration constants of $G=6.1~\AA$/Hz and $G=9.0~\AA$/Hz. The insets display $\Delta\lambda(T)$ from above $T_c$ down to the base temperature. In $\mathrm{Lu}_3\mathrm{Os}_4\mathrm{Ge}_{13}$  a sharp superconducting transition with $T_c \simeq$ 3.05~K is observed, which is the same value as obtained from the ac magnetic susceptibility, while $\mathrm{Y}_3\mathrm{Ru}_4\mathrm{Ge}_{13}$ has a  superconducting transition with midpoint $T_c \simeq$ 2.4~K, slightly higher than the  ac susceptibility.

In the main panels of Figs.~\ref{fig2} (a) and (b), the solid and dashed lines show the fits to a nodeless $s$-wave model and a model with point nodes ($\sim T^2$), respectively. It is clear that the point node model can not describe the data, nor is there a linear temperature dependence as expected in the case of line nodes in the gap. Instead the data flattens at low temperatures which is not expected for nodal superconductivity, but indicates a fully open superconducting gap. For isotropic $s$-wave superconductors at $T<< T_c$, $\Delta\lambda(T) \varpropto \sqrt{\frac{\pi\Delta(0)}{2k_BT}}e^{-\frac{\Delta(0)}{k_BT}}$,
where $\Delta(0)$ is the gap magnitude at zero temperature. As shown in Fig \ref{fig2} (a), the experimental data of $\mathrm{Lu}_3\mathrm{Os}_4\mathrm{Ge}_{13}$ are well fitted in the low-temperature limit with an energy gap of $\Delta(0)=1.35k_BT_c$. Fitting the data of $\mathrm{Y}_3\mathrm{Ru}_4\mathrm{Ge}_{13}$ gives a gap value of $\Delta(0)=1.5k_BT_c$ as shown in Fig.~\ref{fig2} (b). These values are smaller than the BCS value of $1.76k_BT_c$ expected for an isotropic, weakly-coupled BCS superconductor,  consistent with either gap anisotropy or multiband superconductivity.

\begin{figure}[t]
\includegraphics[width=0.99\columnwidth]{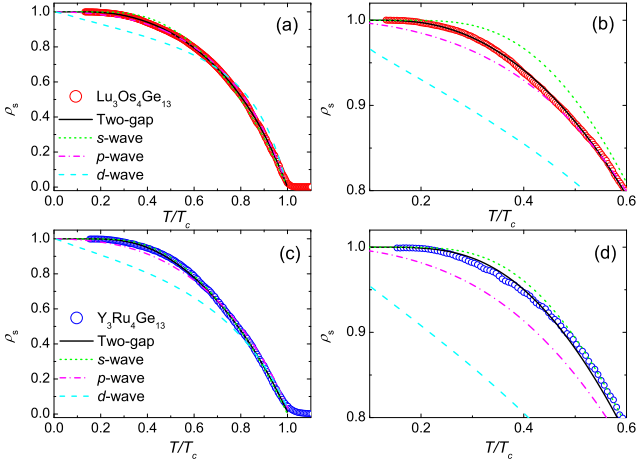}
\caption{(Color online) Normalized superfluid density $\rho_{s}(T)$ along with fits to various models for (a) $\mathrm{Lu}_3\mathrm{Os}_4\mathrm{Ge}_{13}$, where (b) displays an enlargement of the low temperature region, and (c) $\mathrm{Y}_3\mathrm{Ru}_4\mathrm{Ge}_{13}$, with (d) showing the low temperature behavior.}
\label{fig3}
\end{figure}

\begin{figure}[t]
\includegraphics[width=0.99\columnwidth]{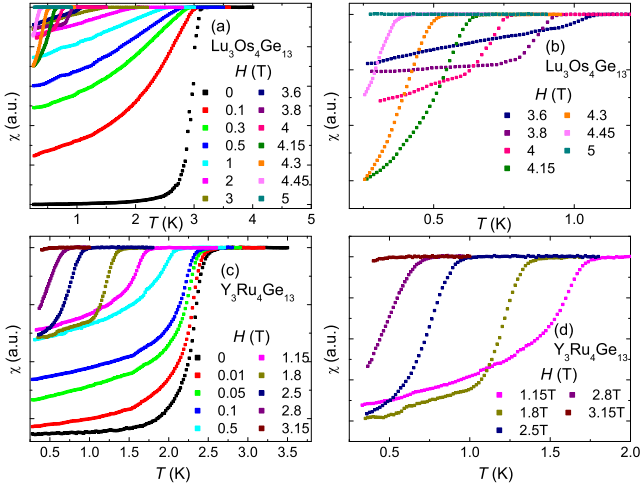}
\caption{(Color online) (a) Temperature dependence of the ac susceptibility of $\mathrm{Lu}_3\mathrm{Os}_4\mathrm{Ge}_{13}$ in various applied magnetic fields, where (b) shows an enlargement of the data for high applied fields.  (c) Temperature dependence of the ac susceptibility of $\mathrm{Y}_3\mathrm{Ru}_4\mathrm{Ge}_{13}$  in various applied magnetic fields, where (d) shows the high field data.}
\label{fig4}
\end{figure}

To obtain more information about the gap structure of $\mathrm{Lu}_3\mathrm{Os}_4\mathrm{Ge}_{13}$ and $\mathrm{Y}_3\mathrm{Ru}_4\mathrm{Ge}_{13}$, the normalized superfluid density $\rho_s(T)$ was calculated from the London penetration depth using $\rho_s(T)=[\lambda(0)/\lambda(T)]^2$, where $\lambda(0)=4736\AA$ \cite{0953-2048-28-11-115012} and 4951$\AA$ \cite{Prakash201390} are calculated from the critical fields of $\mathrm{Lu}_3\mathrm{Os}_4\mathrm{Ge}_{13}$ and $\mathrm{Y}_3\mathrm{Ru}_4\mathrm{Ge}_{13}$ respectively.
The respective $\rho_s(T)$ of $\mathrm{Lu}_3\mathrm{Os}_4\mathrm{Ge}_{13}$ and  $\mathrm{Y}_3\mathrm{Ru}_4\mathrm{Ge}_{13}$  are shown in Figs~\ref{fig3}~(a) and (c).   The normalized superfluid density is calculated using 

\begin{equation}
\rho_{\rm s}(T) = 1 + 2 \left\langle\int_{\Delta_k}^{\infty}\frac{E{\rm d}E}{\sqrt{E^2-\Delta_k^2}}\frac{\partial f}{\partial E}\right\rangle_{\rm FS},
\end{equation}
\label{eq2}
\noindent where $f$ is the Fermi function and $\left\langle\ldots\right\rangle_{\rm FS}$  denotes the average over the Fermi surface. The superconducting gap $\Delta_k(T,\theta,\varphi)=g_k(\theta,\varphi)\Delta(T)$ has an angular dependence $g_k(\theta,\varphi)$ and a temperature dependence given by \cite{Carrington2003205}

\begin{equation}
\begin{aligned}
\Delta(T)=\Delta(0){\rm tanh}\left\{1.82\left[1.018\left(T_c/T-1\right)\right]^{0.51}\right\}.
\label{eq1}
\end{aligned}
\end{equation}

\noindent The flat behavior of the data at low temperatures for both materials clearly deviates from the behavior of nodal gap functions, i.e., a $d$-wave model (line nodes) $g_k(\theta,\varphi)=|\textrm{cos}(2\varphi)|$ and a $p$-wave model (point nodes) $g_k(\theta,\varphi)=|\textrm{sin}(\theta)|$. The data were also fitted using a single band $s$-wave model, where the fitted energy gaps are  $\Delta(0)=2.13k_BT_c$ and  $\Delta(0)=2.05k_BT_c$ for $\mathrm{Lu}_3\mathrm{Os}_4\mathrm{Ge}_{13}$ and $\mathrm{Y}_3\mathrm{Ru}_4\mathrm{Ge}_{13}$ respectively, both larger than the isotropic BCS value of  $1.76k_BT_c$. It can be seen in Figs~\ref{fig3}~(a) and (c)  that this can describe the data at higher temperatures. However, as shown in the low temperature enlargements in Figs~\ref{fig3}~(b) and (d), at low temperatures the data drops more rapidly than expected for the single band models. In particular, for $\mathrm{Lu}_3\mathrm{Os}_4\mathrm{Ge}_{13}$ there is a significant deviation below $T/T_c\approx0.6$, while the difference in the case of $\mathrm{Y}_3\mathrm{Ru}_4\mathrm{Ge}_{13}$  is smaller, below $T/T_c\approx0.5$. Such a low temperature deviation from the fitted single gap model would be expected, since considerably smaller gap  values are obtained from fitting the low temperature  $\Delta\lambda(T)$. These results suggest the presence of multiple energy scales and multiband superconductivity, which is also consistent with the complex Fermi surface revealed by band structure calculations.  \cite{0953-2048-28-11-115012} Therefore the data were fitted using a two gap model where the superfluid density is given by $\rho_s(T)=~x\rho_1(T)+(1-x)\rho_2(T)$, where $\rho_i(T)$ is the superfluid density corresponding to a gap $\Delta_i$ and $x$ is the weight of the contribution from $\Delta_1$.

The data for both materials are well described by such a two gap model, with fitted parameters of $\Delta_1(0)=1.25k_BT_c$, $\Delta_2(0)=2.5k_BT_c$ and $x=0.22$ for $\mathrm{Lu}_3\mathrm{Os}_4\mathrm{Ge}_{13}$, and $\Delta_1(0)=1.4k_BT_c$, $\Delta_2(0)=2.15k_BT_c$ and $x=0.15$  for $\mathrm{Y}_3\mathrm{Ru}_4\mathrm{Ge}_{13}$. In both cases  the values of the
smaller gap are close to those obtained from the low-temperature fit of the penetration depth shown in Fig.~\ref{fig2}, as often found for two-band superconductors \cite{0953-2048-26-4-043001}.  Therefore the penetration depth measurements and the derived superfluid density $\rho_s(T)$ are all consistent with multiband superconductivity in $\mathrm{Lu}_3\mathrm{Os}_4\mathrm{Ge}_{13}$ and $\mathrm{Y}_3\mathrm{Ru}_4\mathrm{Ge}_{13}$, as also suggested from specific heat measurements \cite{0953-2048-28-11-115012,Prakash201390}. It should be noted that the superfluid density of both compounds can also be fitted with an anisotropic $s$-wave model and it is often difficult to distinguish between this scenario and two-gap superconductivity from thermodynamic measurements. \cite{0953-2048-26-4-043001} However the  three-dimensional cubic crystal structure and the presence of multiple bands crossing the Fermi level, favors the two-gap scenario.

\subsection{$H-T$ phase diagram and the peak effect}

\begin{figure}[t]
\includegraphics[width=0.4\textwidth]{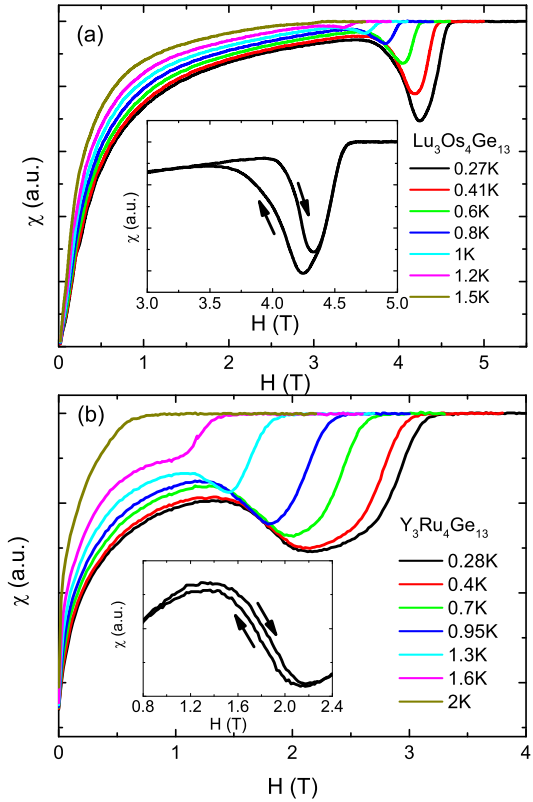}
\caption{(Color online) (a) Isothermal ac susceptibility of $\mathrm{Lu}_3\mathrm{Os}_4\mathrm{Ge}_{13}$ at different temperatures. The inset shows the magnified view of the hysteresis between up and down sweeps of the magnetic field at $T=0.27$~K. (b) Isothermal ac susceptibility of $\mathrm{Y}_3\mathrm{Ru}_4\mathrm{Ge}_{13}$ at different temperatures. The inset shows the magnified view of the hysteresis between up and down sweeps of the magnetic field at $T=0.28$~K.}
\label{fig5}
\end{figure}

\begin{figure}[t]
\includegraphics[width=0.49\textwidth]{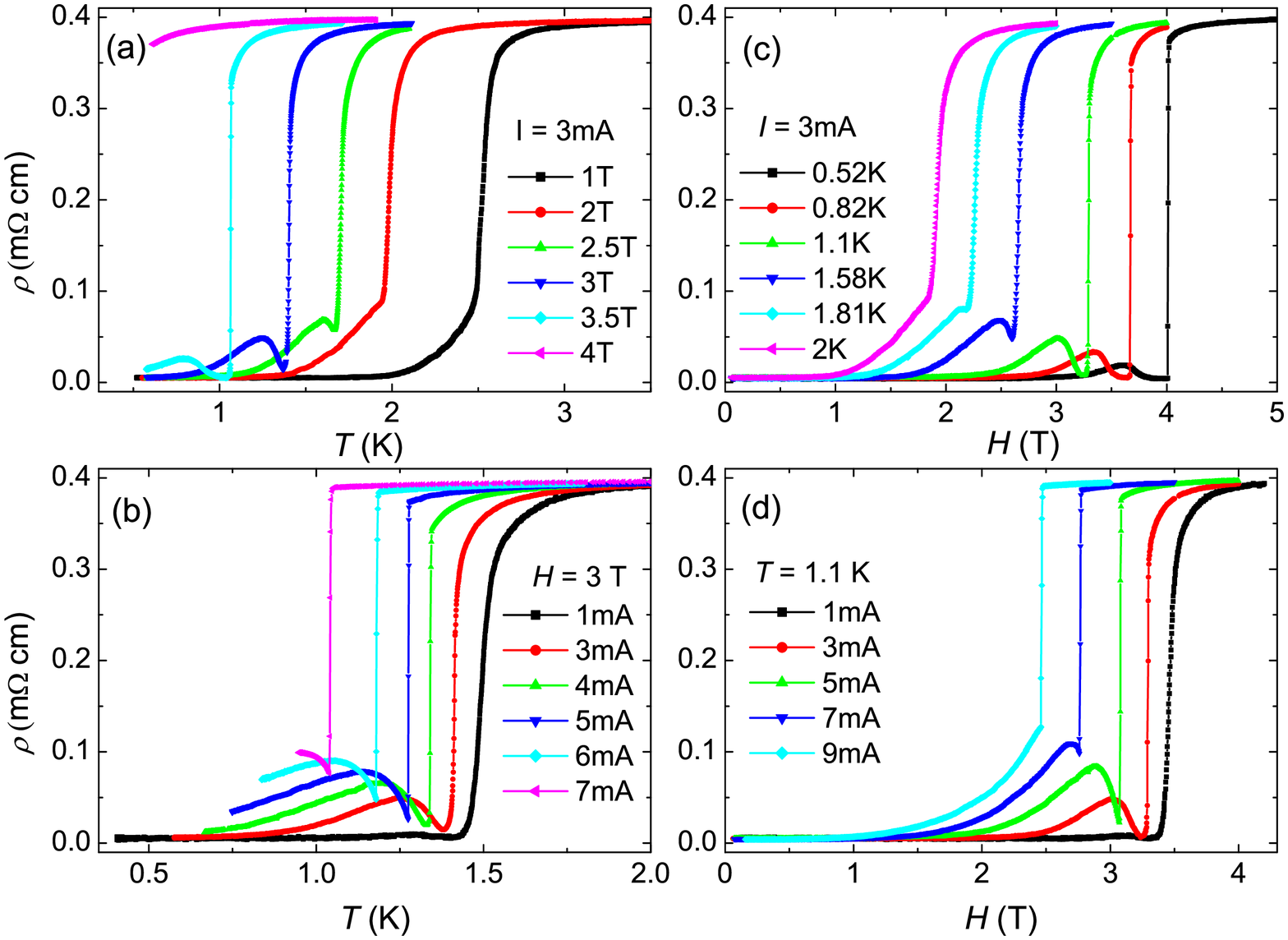}
\caption{(Color online) Resistivity measurements of $\mathrm{Lu}_3\mathrm{Os}_4\mathrm{Ge}_{13}$. The resistivity as  a function of temperature is shown (a) for  various applied fields for a current of $I=3$~mA, and (b) for various currents when 3~T is applied. The field dependence of the resistivity is displayed (c)  at various temperatures with $I=3$~mA, and (d) for various currents at 1.1~K.}
\label{fig6}
\end{figure}

In order to determine the field-temperature phase diagram and probe the properties in the mixed state of $\mathrm{Lu}_3\mathrm{Os}_4\mathrm{Ge}_{13}$ and $\mathrm{Y}_3\mathrm{Ru}_4\mathrm{Ge}_{13}$, isofield and isothermal ac susceptibility measurements were performed. Figure~\ref{fig4}~(a) displays the temperature dependence of the ac susceptibility of $\mathrm{Lu}_3\mathrm{Os}_4\mathrm{Ge}_{13}$ in  various applied magnetic fields, where the sample was cooled in zero field before data were collected upon warming to above $T_c$. In zero field, a sharp superconducting transition is observed but upon
 increasing the field in the mixed state, the transition becomes broader and the shielding fraction is reduced. However, as shown in Fig.~\ref{fig4}~(b), when a field of 3.8~T is applied, the ac susceptibility is reduced at the lowest temperatures compared to 3.6~T, indicating an increase of the superconducting shielding fraction, although $T_c$ continues to decrease. The shielding fraction continues to increase with increasing field until 4.15~T, above which the shielding is again reduced and at 5~T no superconducting transition is observed in the ac susceptibility. Similar behavior is also found in  $\mathrm{Y}_3\mathrm{Ru}_4\mathrm{Ge}_{13}$ (Figs.~\ref{fig4}(c) and (d)), where the shielding fraction again begins to increase above 1.15~T, before again decreasing between 1.8 and 2.5~T.

The field dependence of the ac susceptibility is also displayed for $\mathrm{Lu}_3\mathrm{Os}_4\mathrm{Ge}_{13}$ in Fig.~\ref{fig5}~(a) and $\mathrm{Y}_3\mathrm{Ru}_4\mathrm{Ge}_{13}$ in Fig.~\ref{fig5}~(b). The sample was first cooled to a given temperature and the data were then collected upon increasing and decreasing the field and for clarity, the main panels only show the down sweeping curves. For $\mathrm{Lu}_3\mathrm{Os}_4\mathrm{Ge}_{13}$ at 0.27~K, $\chi(H)$  increases with increasing field at low fields, reaching a peak at around 3.5~T before decreasing to a minimum at 4.25~T. This minimum lies  below the upper critical field $H_{c2}$, above which $\chi(H)$ flattens. As displayed in the inset, the  curves measured with increasing and decreasing field split in the region where  $\chi(H)$ has a negative slope, whereas $\chi(H)$ is reversible at lower fields. With increasing temperature, the magnitude of the decrease  above the peak reduces  and at around 1.5~K, the anomaly is  barely resolvable.  The size of the hysteresis in the vicinity of the anomaly also decreases with increasing temperature. A similar anomaly is also seen in  $\mathrm{Y}_3\mathrm{Ru}_4\mathrm{Ge}_{13}$, where the  local minimum is present at around  2.8~T at 0.28~K. Compared to  $\mathrm{Lu}_3\mathrm{Os}_4\mathrm{Ge}_{13}$, the magnitude of the dip above the peak is more shallow and as displayed in the inset, the  hysteresis is also reduced. The ac susceptibility measurements indicate the presence of the peak effect in both compounds, where in a certain field range near $H_{c2}$, there is an increase of the critical current $J_c$ with increasing field instead of a decrease, which also accounts for the  greater hysteresis between the up-sweeping and down-sweeping measurements. 

\begin{figure}[tb]
\includegraphics[width=0.351\textwidth]{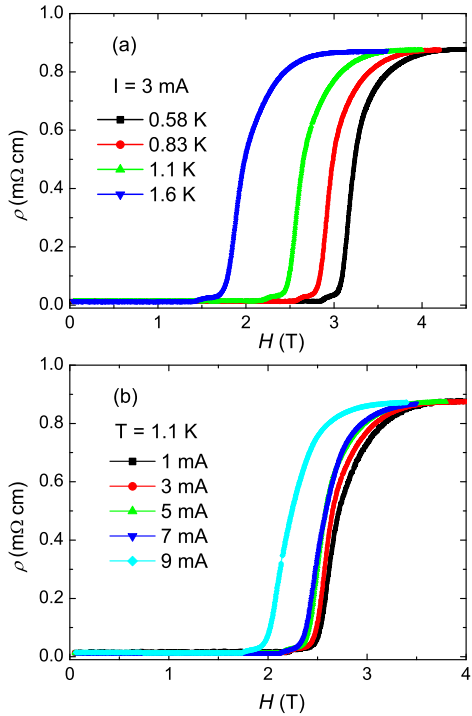}
\caption{(Color online) Resistivity measurements of $\mathrm{Y}_3\mathrm{Ru}_4\mathrm{Ge}_{13}$, (a) as a function of field at various temperatures with a current of $I=3$~mA and (b) as a function of field at 1.1~K with various currents.}
\label{fig8}
\end{figure}

\begin{figure}[t]
\includegraphics[width=0.4\textwidth]{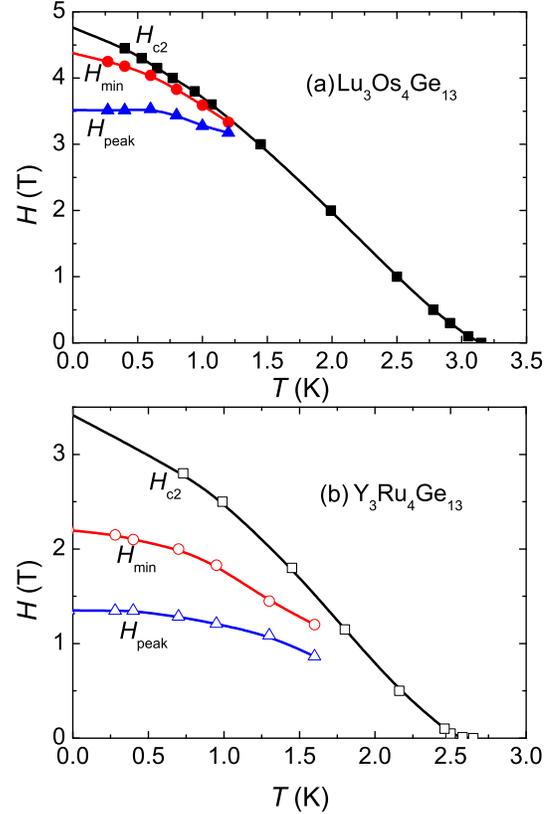}
\caption{(Color online) Field-temperature phase diagram of  (a) $\mathrm{Lu}_3\mathrm{Os}_4\mathrm{Ge}_{13}$, and  (b) $\mathrm{Y}_3\mathrm{Ru}_4\mathrm{Ge}_{13}$. The black squares display the values of the upper critical field $H_{c2}$ obtained from the temperature dependence of the ac susceptibility, while the red circles and blue triangles display the positions of the minima ($H_{\rm min}$) and maxima ($H_{\rm peak}$) in the field dependence of the ac susceptibility.}
\label{fig7}
\end{figure}

Figure~\ref{fig6}~(a) shows the temperature dependence of the resistivity of $\mathrm{Lu}_3\mathrm{Os}_4\mathrm{Ge}_{13}$ under various magnetic fields with a current of $I=3$~mA. In applied magnetic fields of 1~T and 2~T, after a sharp drop at the transition, there is a broad tail at lower temperatures where the resistivity is non-zero. At larger magnetic fields in  the range of 2.5~-~3.5~T, a peak in the resistivity appears below $T_c$. When 3.5~T is applied it can be clearly seen that upon cooling through the transition, zero resistivity is reached but when the sample is cooled further  the resistivity becomes finite, indicating the presence of dissipative processes in the superconducting state, which disappear close to $H_{c2}$ at high fields. At 4~T only a very weak downturn is observed despite the clear transition in the ac susceptibility for these fields, but this could also arise due to self heating as a result of the fairly large current. The temperature dependence of the resistivity is shown for different currents in an applied field of 3~T in Fig.~\ref{fig6}~(b). For larger currents, the transition is seen at lower temperatures, which is again likely due to self heating. At 1~mA, a significant finite resistivity is not observed below the transition, but at higher currents a peak is observed with the maximum resistivity increasing with increasing current. This suggests that the finite resistivity arises in the superconducting state due to the current inducing a  sufficiently large Lorentz force to cause movement of vortices and therefore leading to dissipation. The unpinning of vortices within the superconducting state was also inferred from magnetization measurements at low fields. \cite{0953-2048-28-11-115012}   The field dependence of the resistivity is shown for  $I=3$~mA at various temperatures in Fig.~\ref{fig6}~(c). Upon increasing the field at 0.52~K, the resistivity remains zero up to around 2.8~T, before it begins to increase, reaching a maximum at about 3.6~T. At higher fields the resistivity decreases, reaching zero again at around 3.9~T and it remains zero until a sharp jump to the normal state at about 4~T. At higher temperatures, the peak moves to lower field and is clearly seen up to at least 1.58~K. At higher temperatures still, a finite resistivity is still found in the superconducting state, but without a clear peak. For different currents at a fixed temperature [Fig.~\ref{fig6}~(d)], the peak moves to lower field with increasing current until it can not be clearly resolved. 

In contrast, for  $\mathrm{Y}_3\mathrm{Ru}_4\mathrm{Ge}_{13}$ a finite resistivity is not observed at low fields  and as displayed in Fig.~\ref{fig8}, for all measured currents and temperatures the resistivity only becomes non-zero just below the transition to the normal state at $H_{c2}$. Therefore  in the low field region, the Lorentz force is not sufficiently strong to unpin vortices and lead to significant current dissipation, but just below  $H_{c2}$ the pinning is weak enough for vortices to move.

The field-temperature phase diagrams for both materials are displayed in Fig.~\ref{fig7}, where $H_{c2}$ is determined from the onset of the transition in the ac susceptibility, while  $H_{\rm min}$ and $H_{\rm peak}$ correspond to the minima and maxima of the field dependent ac susceptibility, respectively. Extrapolating to zero temperature gives respective values of $H_{c2}(0)$ of 4.8~T and 3.4~T for $\mathrm{Lu}_3\mathrm{Os}_4\mathrm{Ge}_{13}$ and $\mathrm{Y}_3\mathrm{Ru}_4\mathrm{Ge}_{13}$. In addition, $H_{c2}(T)$ of both compounds shows an upturn with decreasing temperature near $T_c$. This behavior has  often been observed in multiband superconductors, \cite{0953-2048-26-4-043001} and  is therefore further evidence for two-gap superconductivity. The observation of the peak effect indicates that at low temperatures, there is a region of the phase diagram where $J_c$ increases with increasing field, indicating a change from a more weakly pinned region at low fields to stronger pinning at high fields. In $\mathrm{Lu}_3\mathrm{Os}_4\mathrm{Ge}_{13}$ the pinning at low fields is sufficiently weak that the resistivity becomes non-zero in the mixed state, but the stronger pinning at high fields leads to the peak features observed in Figure~\ref{fig6}, whereas in $\mathrm{Y}_3\mathrm{Ru}_4\mathrm{Ge}_{13}$ the flux lines remain static in the region with weaker pinning at low fields. The stronger pinning in $\mathrm{Y}_3\mathrm{Ru}_4\mathrm{Ge}_{13}$ compared to $\mathrm{Lu}_3\mathrm{Os}_4\mathrm{Ge}_{13}$ is consistent with a larger  normal state resistivity, indicating greater disorder. The occurrence of the peak effect has been explained as arising due to the rigidity of the flux line lattice disappearing with field more rapidly than the pinning force. \cite{Pippard1969} Peak effect behavior has been observed in various superconductors such as CeRu$_2$ \cite{CeRu2PE1,CeRu2PE2}, 2H-NbSe$_2$ \cite{NbSe2PE1,NbSe2PE2} and V$_3$Si \cite{V3SiPE1}. It is also  a common phenomenon in $R_3T_4\mathrm{Sn}_{13}$ materials, having been observed in   $\mathrm{Yb}_3\mathrm{Rh}_4\mathrm{Sn}_{13}$\cite{doi:10.1143/JPSJ.64.3175,Tomy19971}, $\mathrm{Ca}_3\mathrm{Rh}_4\mathrm{Sn}_{13}$ \cite{PhysRevB.56.8346} and $\mathrm{Ca}_3\mathrm{Ir}_4\mathrm{Sn}_{13}$, \cite{Kumar201469} and this may correspond to a change from a well ordered vortex lattice, to a disordered or partially ordered vortex glass phase. \cite{Kumar201469,PhysRevB.61.12394,Sarkar2001,PhysRevB.90.020507,Mazzone2015}

\section{summary}
In summary, we have performed measurements of the London penetration depth $\Delta\lambda$(T) of $\mathrm{Lu}_3\mathrm{Os}_4\mathrm{Ge}_{13}$  and $\mathrm{Y}_3\mathrm{Ru}_4\mathrm{Ge}_{13}$ single crystals using a TDO-based technique down to 0.4 K. In both materials the behavior of $\Delta\lambda(T)$ at low temperatures  clearly indicates nodeless superconductivity, while the analysis of $\Delta\lambda(T)$ and the superfluid density $\rho_s(T)$ gives evidence for the presence of multiple gaps. We also mapped the field-temperature phase diagram and at low temperatures we find  the peak effect in the ac susceptibility, which corresponds to an increase of $J_c$ with field, as also reported in some of the isostructural stannides. Further work is required to understand the arrangement of vortices in the mixed state of these materials, and whether this behavior leads to the disordering of the vortex lattice at high fields, which require further measurements such as small angle neutron scattering.

\begin{acknowledgments}
We thank F. Steglich, Q. H. Chen and X. Lu for valuable discussions.This work was supported by the National Natural Science Foundation of China (U1632275, No. 11474251), National Key Research and Development Program of China (No. 2016YFA0300202), and the Science Challenge Project of China.
\end{acknowledgments}

\end{document}